\documentclass[aip,pop,reprint]{revtex4-1}

\usepackage{amsmath}
\usepackage{graphicx}
\usepackage{bm}

\begin{document}

\title{Effect of collisions on the two-stream instability in a finite length plasma}

\author{D. Sydorenko}

\affiliation{University of Alberta, Edmonton, Alberta T6G 2E1, Canada}

\author{I. D. Kaganovich}

\affiliation{Princeton Plasma Physics Laboratory, Princeton University,
          Princeton, New Jersey 08543, USA}

\author{P. L. G. Ventzek}

\author{L. Chen}

\affiliation{Tokyo Electron America, Austin, Texas 78741, USA}

\begin{abstract}

The instability of a monoenergetic electron beam in a collisional
one-dimensional plasma bounded between grounded walls is considered both
analytically and numerically. Collisions between electrons and neutrals are
accounted for the plasma electrons only. Solution of a dispersion equation
shows that the temporal growth rate of the instability is a decreasing linear
function of the collision frequency which becomes zero when the collision
frequency is two times the collisionless growth rate. This result is
confirmed by fluid simulations. Practical formulas are given for the estimate
of the threshold beam current which is required for the two-stream
instability to develop for a given system length, neutral gas pressure,
plasma density, and beam energy. Particle-in-cell simulations carried out
with different neutral densities and beam currents demonstrate good agreement
with the fluid theory predictions for both the growth rate and the threshold
beam current.

\end{abstract}




\maketitle

\section{Introduction \label{sec:01}}

In plasma discharges, surfaces of objects immersed in the plasma may emit
electrons. This emission may be caused by direct heating of the material
(thermal cathodes), intense light (photoemission), or by bombardment by
energetic electrons or ions (secondary electron emission). If the electrostatic
potential of the object is much lower than the potential of the plasma, the
emitted electrons are accelerated by the electric field and form an electron
beam in the plasma. If the two-stream instability develops, the intense plasma
oscillations will heat the plasma and modify the electron velocity distribution
function (EVDF). This includes, in particular, production of suprathermal
electrons which are very important in material processing.~\cite{XuAPL2008}

The presence of the electron beam itself does not guarantee that the intense
oscillations will appear. If the dissipation, \textit{e.g.} due to collisions
between electrons and neutrals, is strong, the instability may not develop at
all. In an experiment of Sato and Heider,\cite{SatoJPDAP1975} a 1000 eV
electron beam was creating a plasma in a chamber filled with neutral gas
(hydrogen or helium). Various values of neutral pressure were used. An energy
analyzer placed at the exit end was measuring the energy spectrum of the beam
particles. For neutral pressures below 70 mTorr, it was found that the higher
the beam current the greater the width of the spectrum. Similar effect was
caused by the increase of the neutral density since this was accompanied with
the plasma density increase and the corresponding increase of the instability
growth rate. The growth of the energy spectrum width was attributed to the
nonlinear interactions between the beam and the plasma. However, when the
neutral pressure exceeded 70 mTorr, the spectrum broadening disappeared which
was interpreted as suppression of the instability by the collisions.

The effect of collisions of plasma electrons on the two-stream instability has
been studied previously in infinite or semi-infinite plasmas.
Singhaus\cite{SinghausPF1964} considered a relativistic electron beam with a
Gaussian velocity distribution in a relatively cold plasma. The system was
infinite and initially uniform. Only collisions for plasma electrons were
accounted for. Electron beams with different temperatures were used. The beam
was considered hot or cold depending on the value of parameter
$\tau=({n_p}{T_b}/{n_b}{W_b})^{1/2}$, where $n_b$ and $n_p$ are the beam and
the plasma electron densities, and $W_b$ and $T_b$ are the average energy and
temperature of the beam.
For the cold beam $\tau\ll 1$ and low collision frequency $\nu\ll Im(\omega)$,
the growth rate is
\begin{equation}\label{eq:s2}
  Im(\omega)=0.69(\omega_p\omega_b^2)^{1/3}~,
\end{equation}
where $\omega_p$ and $\omega_b$ are the Langmuir frequencies of the plasma and
the beam electrons. For the large collision frequency $\nu\gg Im(\omega)$, the
growth rate of the cold beam instability is
\begin{equation}\label{eq:s3}
  Im(\omega)=0.8\left(\omega_p\omega_b^2/\nu\right)^{1/2}~.
\end{equation}
Growth rate (\ref{eq:s2}) is essentially an ordinary collisionless growth rate
of the two-stream instability in an infinite plasma. Growth rate (\ref{eq:s3})
remains nonzero for any value of $\nu$ and therefore the collisions cannot
stabilize the cold beam. The reason for this is the strong resonance between
the cold beam and the plasma wave in the infinite plasma. In warm beams,
however, this resonance is weaker and the collisions may prevent the
instability. According to Singhaus,\cite{SinghausPF1964} for high-temperature
beams $\tau\gg 1$ this occurs if the collision frequency satisfies condition
\begin{equation}\label{eq:s1}
  \nu\gtrsim 0.76\omega_p\tau^{-2}~.
\end{equation}
These conclusions, with minor corrections, were later confirmed by Self,
Shoucri, and Crawford.~\cite{SelfJAP1971} In particular, they provide the
following convenient expression for the growth rate of the instability of a hot
Maxwellian beam consistent with the criterion (\ref{eq:s1}):
\begin{equation}\label{eq:s4}
  Im(\omega)=0.38\omega_p\tau^{-2}-\nu/2~.
\end{equation}

Abe et al.~\cite{AbePF1979} used a one-dimensional particle-in-cell (PIC) code
to study injection of a beam into a long (15 to 17 resonance beam wavelengths
$\lambda_b=2\pi v_b/\omega_e$ where $v_b$ is the beam velocity and $\omega_e$
is the electron plasma frequency) plasma-filled region approximating a
semi-infinite plasma. They focused on spatial rather than temporal growth and
found that while the linear spatial growth rate (observed in the region few
$\lambda_b$ wide near the emission boundary) was barely affected by collisions,
the amplitude of oscillations in the area where the particle dynamics is
nonlinear (downstream of the first wave amplitude maximum) was reduced compared
to the collisionless case. It is necessary to mention that the collisions
considered were due to electrostatic forces existing in the numerical model. In
general, the frequency of such collisions is a function of the grid size, time
step, and the number of particles.\cite{DawsonPF1964,LewisJCP1972}

Andriyash, Bychenkov, and Rozmus\cite{AndriyashHEDP2008} considered
analytically and numerically the two-stream instability which appears when an
ultra-short linearly-polarized X-ray laser pulse produces streams of
photoionized electrons in a gas target. They shown that for higher collision
frequencies, the saturation of the instability occurs faster and at lower
levels.

Cottrill et al.\cite{CottrillPP2008} considered relativistic electron beams
with different temperatures and distribution functions propagating in a dense
plasma where electron-ion colisions for the bulk electrons are important. This
problem is relevant to the fast ignition scheme in fusion applications where
the plasma is heated by a relativistic beam created by a short high-intensity
laser pulse. Numerical solution of the dispersion equation showed that the
two-stream instability growth rate is reduced due to collisions for the cold
beam. For a high-temperature beam the collisions can cancel the two-stream
instability. Similar results were obtained by Hao et al.~\cite{HaoPRE2009} who
also solved a kinetic dispersion relation for a relativistic electron beam
propagating in a cold dense plasma with Coulomb collisions.

Lesur and Idomura\cite{LesurNF2012} studied the bump-on-tail instability in an
infinite 1D plasma using a Vlasov code with the collisional operator containing
drag and diffusion. They showed that the collisions strongly affect nonlinear
stochastic dynamics of plasma oscillations in such a system.

Unlike papers mentioned above, the present paper considers a low pressure
beam-plasma system of the length of only few beam resonance wavelengths.
Previously, a short beam-plasma system was considered by
Pierce.\cite{PierceJAP1944} In Pierce diode, however, the beam electron density
is equal to the density of background ions, while in the present paper there is
also the electron background while the beam density is small compared to the
density of ambient electrons.
%
For theoretical analysis, the electron beam is considered as monoenergetic.
This is justified if the beam energy is much higher than both the plasma and
the beam temperature. 
The beam
temperature may increase as the beam propagates through the plasma due to
scattering on neutrals and due to interaction with strong plasma wave if the
two-stream instability is excited. However, the size of the system considered
is small compared to the mean free path of beam electrons associated with
electron-neutral collisions. Beam electrons perturbed by the wave also quickly
leave the system and perturbations of the beam velocity remain small compared
to the initial velocity for a large part of the linear growth stage.
Whether the two-stream instability develops or not depends on the competition
between the two time scales -- the growth rate of the instability without
collisions and the collision frequency.
The growth rate of the instability in such a short plasma is much lower than
that in an infinite plasma.\cite{KaganovichArxiv2015}
Therefore, the suppression of the instability occurs for a lower neutral gas
pressures than in an infinite plasma for the same beam and plasma density and
the beam energy.



The paper is organized as follows. In Section~\ref{sec:02}, the linear theory
of the two-stream instability in a bounded plasma with collisions is given. In
Section~\ref{sec:03}, fluid simulation confirms the theory. Kinetic simulations
are described in Section~\ref{sec:04}. The results are summarized in
Section~\ref{sec:05}.

\section{Fluid theory \label{sec:02}}

Consider a cold uniform plasma bounded between two grounded walls at $x=0$ and
$x=L$. A beam of electrons is emitted by the wall $x=0$ with velocity $v_{b,0}$
and absorbed by the wall $x=L$. The walls reflect plasma electrons. This
boundary condition approximates a sheath appearing in a real plasma at the
plasma-wall interface. The beam motion is collisionless while the plasma
electrons are scattered, for example by neutrals, with the collision frequency
$\nu_e$. The ion motion is omitted. The ion density $n_i$ is uniform, constant,
and it ensures that the plasma is initially neutral $n_i=n_{e,0}+n_{b,0}$,
where $n_{e,0}$ and $n_{b,0}$ are the initial densities of the bulk and beam
electrons. Full dynamics of such a system is described by the following set of
equations:
\begin{equation}\label{eq:01}
    \frac{\partial}{\partial t}n_{e,b}+
    \frac{\partial}{\partial x}n_{e,b}v_{e,b}=0~,
\end{equation}
\begin{equation}\label{eq:02}
    \frac{\partial}{\partial t}v_e+
    v_e\frac{\partial}{\partial x}v_e=-\frac{e}{m}E-\nu_e v_e~,
\end{equation}
\begin{equation}\label{eq:03}
    \frac{\partial}{\partial t}v_b+
    v_b\frac{\partial}{\partial x}v_b=-\frac{e}{m}E~,
\end{equation}
\begin{equation}\label{eq:04}
    \frac{\partial^2}{\partial x^2}\Phi=\frac{e}{\varepsilon_0}(n_e+n_b-n_i)~,
\end{equation}
where subscripts \textit{e} and \textit{b} denote bulk and beam electrons, $-e$
and $m$ are the electron charge and mass, the electric field is
$E=-\partial\Phi/\partial x$, and $\Phi$ is the electrostatic potential.

Note that in the beam motion equation (\ref{eq:03}) the collisional term is
omitted. This is reasonable if (a) for the given beam energy and neutral
density the electron mean free path is much larger than the size of the system
$L$, and (b) the energy of the beam is high (typically hundreds of eV) so that
the scattering occurs at small angles~\cite{OkhrimovskyyPRE2002,KhrabrovPP2012}
and the velocity of the scattered beam electron is close to its initial
velocity. Collisions for the plasma electrons are retained, however, since
these electrons are trapped by the sheath inside the plasma volume and they
suffer collisions even though they have to bounce several times between the
walls before that. These conditions are satisfied in kinetic simulations
described in Section~\ref{sec:04}.

The linear dispersion equation is obtained in a procedure similar to the one
used by Pierce~\cite{PierceJAP1944} and recently in
Ref.~\onlinecite{KaganovichArxiv2015}. The plasma and the beam densities and
velocities are represented as sums of unperturbed values $n_{e,0}$, $n_{b,0}$,
$v_{b,0}$ and perturbations $\delta n_e$, $\delta n_b$, $\delta v_b$, and
$\delta v_e$ (the unperturbed value of the plasma electron velocity is zero).
The perturbations are described by linearized equations
(\ref{eq:01}-\ref{eq:04}):
\begin{equation}\label{eq:05}
    \frac{\partial}{\partial t}\delta n_e+
    n_{e,0}\frac{\partial}{\partial x}\delta v_e=0~,
\end{equation}
\begin{equation}\label{eq:06}
    \frac{\partial}{\partial t}\delta v_e
    =-\frac{e}{m}E-\nu_e \delta v_e~,
\end{equation}
\begin{equation}\label{eq:07}
    \frac{\partial}{\partial t}\delta n_b+
    \frac{\partial}{\partial x}(n_{b,0}\delta v_b+\delta n_b v_{b,0})=0~,
\end{equation}
\begin{equation}\label{eq:08}
    \frac{\partial}{\partial t}\delta v_b+
    v_{b,0}\frac{\partial}{\partial x}\delta v_b=-\frac{e}{m}E~,
\end{equation}
\begin{equation}\label{eq:09}
    \frac{\partial^2}{\partial x^2}\Phi=
    \frac{e}{\varepsilon_0}(\delta n_e+\delta n_b)~,
\end{equation}
Note that for perturbations proportional to $\exp(-i\omega t + ikx)$, equations
(\ref{eq:05}-\ref{eq:09}) give a usual dispersion equation
\begin{equation}\label{eq:10}
    1=\frac{\omega_{e,0}^2}{\omega(\omega + i\nu_e)}+
    \frac{\omega_{b,0}^2}{(\omega-k v_{b,0})^2}~,
\end{equation}
where $\omega_{e,0}^{2}\equiv n_{e,0}e^2/\varepsilon_0 m$ and
$\omega_{b,0}^{2}\equiv n_{b,0}e^2/\varepsilon_0 m$ are the electron plasma
frequencies of the plasma and the beam electrons, respectively.

In the bounded system one looks for a solution for the potential in the form
\begin{equation}\label{eq:11}
  \Phi(t,x)=\left( {Ax+Be^{ik_{+}x}+Ce^{ik_{-}x}+D}\right)
                   e^{-i\omega t},
\end{equation}
where coefficients $A,B,C,D$ are complex constants and wave vectors $k_{\pm }$
of the two waves propagating in the system satisfy dispersion equation
(\ref{eq:10}):
\begin{equation}\label{eq:12}
    (\omega-k_\pm v_{b,0})^2 =
    \frac{\omega_{b,0}^2}
         {1-{\omega_{e,0}^2}/{\omega(\omega + i\nu_e)}}~.
\end{equation}
The corresponding density and velocity perturbations of plasma and beam
electrons are
\begin{equation}\label{eq:13}
\begin{split}
  \delta {n_{e,b}}(t,x)& =
     \left( \delta {n_{e,b}^{\prime }}+
            \delta {n_{e,b}^{+}}e^{ik_{+}x}+
            \delta {n_{e,b}^{-}}e^{ik_{-}x}\right) e^{-i\omega t},
  \\
  \delta {v_{e,b}}(t,x)& =
     \left( \delta {v_{e,b}^{\prime }}+
            \delta {v_{e,b}^{+}}e^{ik_{+}x}+
            \delta {v_{e,b}^{-}}e^{ik_{-}x}\right) e^{-i\omega t}.
\end{split}
\end{equation}
Substituting (\ref{eq:11}) and (\ref{eq:13}) into
Eqs.~(\ref{eq:05}-\ref{eq:08}) gives
\begin{equation}\label{eq:14}
\begin{split}
    \delta n_e^\prime &= 0,~
    \delta n_e^\pm = \frac{n_{e,0} k_\pm}{\omega}\delta v_e^\pm,~
    \delta v_e^\prime = \frac{e}{m}\frac{A}{(-i\omega+\nu_e)}, \\
    \delta v_e^+ &= -\frac{e}{m}\frac{k_+ B}{(\omega+i\nu_e)},~
    \delta v_e^- = -\frac{e}{m}\frac{k_- C}{(\omega+i\nu_e)},
\end{split}
\end{equation}
and
\begin{equation}\label{eq:15}
\begin{split}
    \delta n_b^\prime &= 0,~
    \delta n_b^\pm = \frac{n_{b,0} k_\pm}{\omega-k_\pm v_{b,0}}\delta v_b^\pm,~
    \delta v_b^\prime = \frac{e}{m}\frac{A}{(-i\omega)}, \\
    \delta v_b^+ &= -\frac{e}{m}\frac{k_+ B}{(\omega-k_+ v_{b,0})},~
    \delta v_b^- = -\frac{e}{m}\frac{k_- C}{(\omega-k_- v_{b,0})}.
\end{split}
\end{equation}

Combining (\ref{eq:11}), (\ref{eq:14}), and (\ref{eq:15}) with boundary
conditions $\delta n_b(0)=0$, $\delta v_b(0)=0$, $\Phi(0)=0$, and $\Phi(L)=0$,
one obtains
\begin{equation}\label{eq:16}
\begin{split}
    k_+^2B + k_-^2C = 0, \\
    \frac{A}{\omega}+
    \frac{e}{m}\frac{ik_+ B}{(\omega-k_+ v_{b,0})}+
    \frac{e}{m}\frac{ik_- C}{(\omega-k_- v_{b,0})}=0, \\
    B+C+D=0, \\
    AL+Be^{ik_+L}+Ce^{ik_-L}+D=0,
\end{split}
\end{equation}
which results in the following dispersion equation:
\begin{equation}\label{eq:17}
    -i\frac{2(1-\chi)}{\chi(1+\chi)}+
    e^{i(1-\chi)L_n}-
    1-
    \frac{(1-\chi^2)}{(1+\chi^2)}
    \left[e^{i(1+\chi)L_n}-1\right]=0~,
\end{equation}
where
\begin{equation}\label{eq:aft17}
L_n\equiv L\omega_{e,0}/v_{b,0}
\end{equation}
and
\begin{equation}\label{eq:18}
  \chi=\frac{\omega_{b,0}/\omega_{e,0}}
            {\sqrt{1-{\omega_{e,0}^2}/{\omega(\omega+i\nu_e)}}}.
\end{equation}
Equations (\ref{eq:17}-\ref{eq:18}) define frequency $\omega$ as a function of
the distance between the walls $L$ while the wavenumbers $k_\pm$ can be
obtained from (\ref{eq:12}) and (\ref{eq:18}) as
\begin{equation}\label{eq:19}
    k_\pm v_{b,0}=(1\mp\chi)\omega_{e,0}.
\end{equation}

Dispersion equation (\ref{eq:17}) exactly matches the dispersion equation for
$\chi$ obtained in Ref.~\onlinecite{KaganovichArxiv2015} for the collisionless
system. Therefore, $\chi$ itself is a universal parameter depending on the
normalized system length (\ref{eq:aft17}) only. The difference, however, is in
the definition (\ref{eq:18}) of variable $\chi$ as a function of $\omega$ which
in the present paper involves the collision frequency. Note that if $\nu_e=0$,
equation (\ref{eq:18}) defines $\chi$ in exactly the same way as
Ref.~\onlinecite{KaganovichArxiv2015}. Introduce the frequency of oscillations
in the collisionless system as
\begin{equation}\label{eq:20}
   \omega_{ncl}^2=\frac{\omega_{e,0}^2}{1-\alpha/\chi^2}~,
\end{equation}
where subscript ``ncl'' stands for ``no collisions'', $\alpha=n_{b,0}/n_{e,0}$
is the relative beam density, and $\chi$ is found from (\ref{eq:17}). Then
(\ref{eq:18}) can be transformed to
\begin{equation}\label{eq:21}
   \omega(\omega+i\nu_e)-\omega_{ncl}^2=0~,
\end{equation}
which in the limit $\nu_e\ll\omega_{e,0}$ gives $\omega=\omega_{ncl}-i\nu_e/2$
so that
\begin{equation}\label{eq:22}
   Im(\omega)=Im(\omega_{ncl})-\nu_e/2~.
\end{equation}
Therefore, the instability will not develop if the collision frequency exceeds
a threshold value
\begin{equation}\label{eq:23}
   \nu_{e,thr}=2Im(\omega_{ncl})~,
\end{equation}
where $Im(\omega_{ncl})$ is the temporal growth rate in the system without
collisions which can be estimated using the approximate formula provided in
Ref.~\onlinecite{KaganovichArxiv2015}:
\begin{equation}\label{eq:24}
    \frac{Im(\omega_{ncl})}{\omega_{e,0}}=
    \frac{\alpha}{13}L_n\ln(L_n)
    \left[1-0.18\cos\left(L_n+\frac{\pi}{2}\right)\right]~.
\end{equation}
It is necessary to mention that Eq.~\ref{eq:22} is similar to Eq.~\ref{eq:s4}
originally obtained in Ref.~\onlinecite{SelfJAP1971} except for the definition
of the growth rate without collisions.

For practical use, criterion (\ref{eq:23}) can be written in a form which
involves the neutral gas pressure and the beam current. To do this, first,
assume a linear relation between the electron collision frequency $\nu_e$ and
the neutral gas pressure $p_n$ for the selected neutral species:
\begin{equation}\label{eq:25}
    \nu_e=\kappa(T_e) p_n~,
\end{equation}
where the coefficient $\kappa(T_e)$ depends on the scattering cross sections
and the electron temperature (or on the EVDF if it is not Maxwellian).
Using (\ref{eq:25}) and replacing $\alpha$ in (\ref{eq:24}) with
$J_{b,thr}/en_{e,0}v_{b,0}$, where $J_{b,thr}$ is the threshold beam current
density, criterion (\ref{eq:23}) can be written as follows:
\begin{equation}\label{eq:26}
    J_{b,thr}=\frac{6.5\kappa(T_e)\sqrt{2e\varepsilon_0 n_{e,0} W_{b,0}}}
             {L_n\ln(L_n)\left[1-0.18\cos\left(L_n+{\pi}/{2}\right)\right]}
        p_n~,
\end{equation}
where the beam energy $W_{b,0}$ is in electronvolts. For a given neutral gas
pressure, the two-stream instability does not develop if the beam current
density is below this threshold.

\section{Fluid simulation \label{sec:03}}

Theoretical predictions of Section~\ref{sec:02} are tested in fluid
simulations. The numerical fluid model solves equations
(\ref{eq:01}-\ref{eq:04}) on a regular grid. The model uses SHASTA
method~\cite{BorisJCP1973} to advance densities in Eqs.~(\ref{eq:01}) and a
simple upwind scheme to advance velocities in Eqs.~(\ref{eq:02}) and
(\ref{eq:03}).
%
In order to include reflection of plasma electrons from the sheath, condition
$v_e=0$ is introduced at the ends of the system $x=0$ and $x=L$. The potential
at the system ends is set to zero, $\Phi(0)=\Phi(L)=0$. The beam injection at
boundary $x=0$ is ensured by conditions $n_b(x=0)=n_{b,0}$ and
$v_b(x=0)=v_{b,0}$. No boundary condition is imposed on the plasma density at
both ends of the system and on the beam density and velocity at the exit end
$x=L$. Initially, the bulk and beam electron densities are uniform, the beam
flow velocity is $v_{b,0}$ everywhere, the bulk electron flow velocity in the
inner nodes has a harmonic perturbation
$v_{e,0}=v_0\sin(x\omega_{e,0}/v_{b,0})$ where the amplitude $v_0$ is very
small, $v_0\ll v_{b,0}$. Previously this model was used to study the dispersion
of oscillations excited by an electron beam in a collisionless finite length
plasma.\cite{KaganovichArxiv2015}

Simulations discussed in this section are carried out with the following common
parameters: the initial plasma electron density $n_{e,0}=2\cdot
10^{17}~\text{m}^{-3}$, the Langmuir plasma frequency corresponding to this
density $\omega_{e,0}=2.523\cdot 10^{10}\text{s}^{-1}$, beam energy
$W_{b,0}=800\text{~eV}$ (the velocity corresponding to this energy is
$v_{b,0}=1.678\cdot 10^7\text{m/s}$), beam-to-plasma density ratio
$\alpha=0.0001$, numerical grid cell size $\Delta x=4.156\cdot
10^{-6}\text{~m}$ which is 1/8 of the Debye length for the electron density as
above and the temperature 2 eV, time step $\Delta t=8.258\cdot
10^{-14}\text{~s}$. The selected values of $\Delta x$ and $\Delta t$ ensure
stability of the SHASTA algorithm for electron flows with velocity $\Delta
x/2\Delta t$ which corresponds to the energy of $1800\text{~eV}$. The classical
resonance beam wavelength is $\lambda_b=2\pi
v_{b,0}/\omega_{e,0}=4.178\text{~mm}$.
%
\begin{figure}[tbp]
\includegraphics {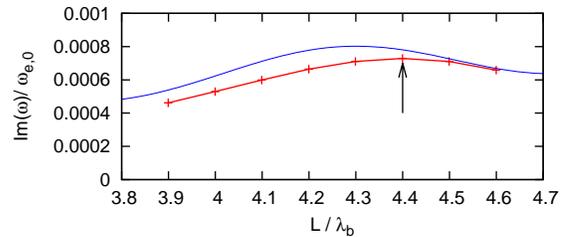}
\caption{\label{fig:ImWvsL_nb0p0001_nocolls} %
Temporal growth rate of the two-stream instability versus system length in a
collisionless beam-plasma system. The red markers connected by the red curve
represent values obtained in the simulation. The solid blue curve is calculated
with the approximate formula (\ref{eq:24}). The arrow marks the length selected
for fluid simulations with non-zero collision frequency shown in
Figures~\ref{fig:logEvst_a0p0001_L4p4} and \ref{fig:ImW_vs_collfreq}. }
\end{figure}

In general, both the frequency and the wavenumber are complex numbers and
functions of $L$ with band structure, as studied in detail in
Ref.~\onlinecite{KaganovichArxiv2015}.
%
The present paper considers a single band with $3.9\le L/\lambda_b\le 4.6$.
This makes the plasma size near 18 mm which is close to the size of the density
plateau in the recent study of beam-plasma interaction in Ref.~\onlinecite{??}
and the size of the plasma in a dc-rf etcher considered in
Ref.~\onlinecite{XuAPL2008}.
The growth rate obtained in collisionless fluid simulation for the selected
band is shown by the red curve in Fig~\ref{fig:ImWvsL_nb0p0001_nocolls}. Note
that there is good agreement between the simulation and the approximate formula
(\ref{eq:24}), compare the red and the blue curves in
Fig.~\ref{fig:ImWvsL_nb0p0001_nocolls}.

\begin{figure}[tbp]
\includegraphics {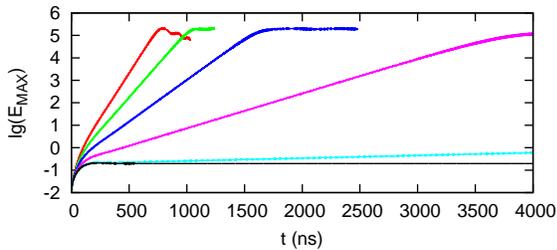}
\caption{\label{fig:logEvst_a0p0001_L4p4} %
Logarithm of the amplitude of electric field in V/m versus time in fluid
simulations with the following relative values of the collision frequency
$\nu/\omega_{e,0}$: 0 (red), $0.0004$ (green), $0.0008$ (blue), $0.00119$
(magenta), $0.00146$ (cyan), and $0.00148$ (black). The dependencies are
obtained at $x=4.178\lambda_b$. The system length is $L=4.4\lambda_b$, see
the arrow in Figure~\ref{fig:ImWvsL_nb0p0001_nocolls}.}
\end{figure}

To study the effect of collisions with the fluid model, the value of
$L=4.4\lambda_b$ is selected corresponding to the maximum of the collisionless
growth rate marked by the arrow in Fig.~\ref{fig:ImWvsL_nb0p0001_nocolls}. A
set of simulations is performed with the frequency of collisions gradually
increasing from zero until the growth of the amplitude of oscillations cancels
completely, as shown in Fig.~\ref{fig:logEvst_a0p0001_L4p4}.
%

The temporal growth rate decreases linearly with the collision frequency in
good agreement with Eq.~\ref{eq:22}, compare the red and the blue curves in
Fig.~\ref{fig:ImWvsL_nb0p0001_nocolls}. Note that the temporal growth rate is
obtained during time intervals of exponential growth with a constant rate
(including the zero growth rate). One can easily identify such intervals in
Fig.~\ref{fig:logEvst_a0p0001_L4p4}. The initial stage when the growth rate
rapidly decreases with time and the saturation stage are excluded from
consideration.
%
\begin{figure}[tbp]
\includegraphics {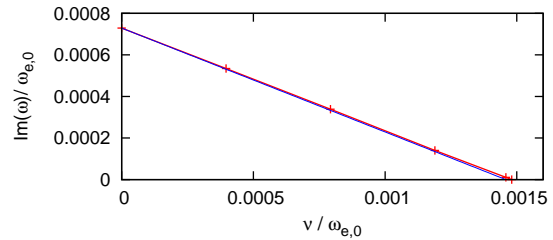}
\caption{\label{fig:ImW_vs_collfreq} %
Temporal growth rate of the two-stream instability versus collision frequency
in a finite-length beam-plasma system. The red markers connected by the red
curve represent values obtained in fluid simulations. The blue curve
corresponds to Eq.~\ref{eq:22}. The system length is $L=4.4\lambda_b$, the
collisionless growth rate is marked by the arrow in
Figure~\ref{fig:ImWvsL_nb0p0001_nocolls}. }
\end{figure}

It is instructive to compare the results presented above with the predictions
of Refs.~\onlinecite{SinghausPF1964,SelfJAP1971}. Equations (\ref{eq:s2}) and
(\ref{eq:s3}) for cold beam are of no use here: the low-collision frequency
limit growth rate (\ref{eq:s2}) is too high and independent on the collision
frequency while the expression for the high-collision frequency growth rate
(\ref{eq:s3}) cannot be used since the collision frequency is too low.
Threshold condition for a warm beam (\ref{eq:s1}), however, produces a
reasonable estimate. The collision frequency which sets the growth rate to zero
in the fluid simulation with $L=4.4\lambda_b$ above is $\nu=3.736\times
10^7\text{s}^{-1}$. For the selected beam and plasma parameters, according to
(\ref{eq:s1}), such a collision frequency cancels the instability in an
infinite plasma if the beam temperature exceeds 41 eV. Similar energy spread of
an 800 eV beam is observed in Ref.~\onlinecite{XuAPL2008}.
Note that using a warm electron beam reduces the growth rate of the two-stream
instability even without collisions, compare equations (\ref{eq:s4}) and
(\ref{eq:s2}). The estimate above shows that for plasma parameters used in the
present paper, which are typical for plasma processing applications, the finite
beam temperature and the finite system length have similar effect for a
reasonable value of the beam temperature. Therefore, in future studies it is
necessary to consider a finite temperature beam in a finite length plasma.

\section{Kinetic simulation \label{sec:04}}

In low-pressure plasmas, kinetic effects are important for the electron
dynamics. The frequency of collisions with neutrals and the scattering angle
for each electron depend on the electron energy. Reflection from the sheath
near the wall changes the direction of the electron velocity but does not stop
the particle. The electron velocity at each point is the result of action of
the electric field not only in this point, but along the whole trajectory
before that. These effects are omitted in the simple fluid approximation used
in Section~\ref{sec:02}. Therefore, it is necessary to check whether the
results of the fluid theory, in particular the expressions for the growth rate
(\ref{eq:22}) and the threshold current (\ref{eq:26}), remain valid if the
kinetic effects are accounted for. Below, for kinetic description of the
interaction of an electron beam with a low-pressure finite length plasma, a
1d3v particle-in-cell (PIC) code EDIPIC~\cite{SydorenkoPhDTh2006} is used.

The PIC simulation setup is very similar to the one in the fluid simulations
above: a uniform plasma is bounded by grounded walls with a distance $L$
between them and wall $x=0$ emits an electron beam with constant flux and
energy. At the beginning of simulation, both beam and bulk electrons are
uniformly distributed along the system. The bulk electrons and the beam
electrons are represented by macroparticles with different charge which greatly
improves resolution of beam dynamics for low beam currents. The ions are
represented by a uniform constant immobile positively charged background which
ensures quasineutrality at $t=0$. The motion of beam electrons is
collisionless, they are injected at $x=0$ with the given beam energy and can
freely penetrate through the boundaries. The bulk electrons may collide
elastically with neutrals and are reflected specularly at the boundaries. The
cross section of the collisions corresponds to Argon.

Initial plasma and beam parameters, such as $n_{e,0}$, $W_b$, and grid
resolution $\Delta x$ are the same as in Section~\ref{sec:03}. The time step is
$\Delta t=1.65\times 10^{-13}\text{~s}$. The plasma electrons have finite
initial temperature of $T_{e,0}=2\text{~eV}$. This value is selected as a
compromise between the desire to have a cold plasma as in the fluid simulation
and the growth of the numerical cost when the Debye length (and therefore the
size of a cell in the computational grid) decreases for low electron
temperature. Note that $T_{e,0}$ is well below the thresholds for the
excitation (11.5 eV) and ionization (15.76 eV) electron-neutral collisions for
Argon which justifies omitting them for the plasma electrons.

\begin{figure}[tbp]
\includegraphics {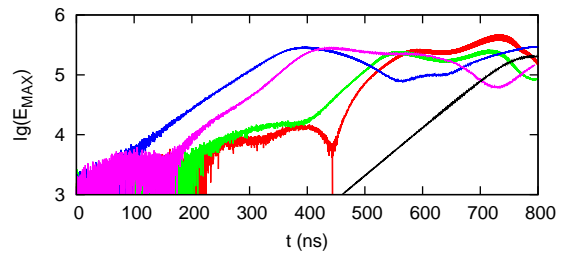}
\caption{\label{fig:pic_logEvst_varL_a0p0001_nocolls} %
Logarithm of the amplitude of electric field in V/m versus time in
collisionless PIC simulations with $\alpha=0.0001$ and the system length
$L=4.3\lambda_b$ (red, obtained at $x=4.16\lambda_b$), $4.4\lambda_b$ (green,
obtained at $x=4.26\lambda_b$), $4.5\lambda_b$ (blue, obtained at
$x=4.31\lambda_b$), and $4.6\lambda_b$ (magenta, obtained at
$x=4.36\lambda_b$). For comparison, the black curve represents a fluid
simulation with $L=4.4\lambda_b$ (same as the red curve in
Figure~\ref{fig:logEvst_a0p0001_L4p4}). Note that the blue curve corresponds to
simulation 5 of Table~\ref{tab:table1}. }
\end{figure}

One unpleasant consequence of using plasma with a finite electron temperature
is that the level of noise in PIC simulations is much higher than that in the
fluid simulation. The noise, in particular, reduces time interval when the
exponential growth of oscillations is visible, see
Fig.~\ref{fig:pic_logEvst_varL_a0p0001_nocolls}.

Kinetic simulation reveals that the beam-plasma system is very sensitive to the
length of the plasma. A set of collisionless simulations is performed with the
relative beam density $\alpha=0.0001$ and $L=4.3\lambda_b$, $4.4\lambda_b$,
$4.5\lambda_b$, and $4.6\lambda_b$. The selected values of $L$ correspond to
the same band of the dispersion as shown in
Fig.~\ref{fig:ImWvsL_nb0p0001_nocolls}. Only for $L=4.5\lambda_b$ the
exponential growth of the amplitude of oscillations has a constant rate from
the start till the first amplitude maximum, see the blue curve in
Fig.~\ref{fig:pic_logEvst_varL_a0p0001_nocolls}. For other values of $L$, the
growth rate oscillates, see the red, green, and magenta curves in
Fig.~\ref{fig:pic_logEvst_varL_a0p0001_nocolls}. Note that simulations with
$L=4.4\lambda_b$ has a noticeable time interval where the growth rate is very
close to the one in the fluid simulation with the same $L$, compare the green
and the black curves in Fig.~\ref{fig:pic_logEvst_varL_a0p0001_nocolls}. It is
necessary to mention that oscillations of the growth rate are observed for
certain intervals of $L$ in fluid simulations as
well,\cite{KaganovichArxiv2015} but these intervals are more narrow. In view of
the above, for PIC simulation with collisions described below the system length
$L=4.5\lambda_b$ is selected.
%

\begin{table}
\caption{\label{tab:table1}%
Parameters of PIC simulations with $L=4.5\lambda_b$. The neutral pressure $p_n$
corresponds to the given neutral gas density $n_n$ and the neutral gas
temperature of $300\text{~K}$. In the growth rate column, ``N/A'' means that
the instability is suppressed and a sufficiently long time interval with
exponential growth cannot be identified. }
\begin{ruledtabular}
\begin{tabular}{ccccccc}
 Number & $\alpha$    & $J_b$ & $n_n$ & $p_n$ & $Im(\omega)$ & $p_{thr}$ \\
        & ($10^{-4}$) & [$\text{A/m}^2$] & [$10^{20}\text{~m}^{-3}$] & [mTorr] & [$10^7\text{~s}^{-1}$] & [mTorr]  \\
\hline
1 & 0.5 & 26.83 & 0.805 & 2.5 & 0.63 & 8.8 \\
2 &     &       & 1.61  & 5   & 0.33 &  \\
3 &     &       & 2.415 & 7.5 & 0.08 &  \\
4 &     &       & 3.22  & 10  & N/A  & \\
\hline
5 & 1   & 53.67 & 0    & 0  & 1.41 & 15.2 \\
6 &     &       & 1.61 & 5  & 1.01 & \\
7 &     &       & 3.22 & 10 & 0.38 & \\
8 &     &       & 4.83 & 15 & N/A  & \\
9 &     &       & 6.44 & 20 & N/A  & \\
\hline
10 & 1.5 & 80.51 & 3.22 & 10 & 1.34 & 25.0 \\
11 &     &       & 4.83 & 15 & 1.01 & \\
12 &     &       & 6.44 & 20 & 0.5 & \\
13 &     &       & 8.05 & 25 & N/A & \\
14 &     &       & 9.66 & 30 & N/A & \\
\end{tabular}
\end{ruledtabular}
\end{table}

%
\begin{figure*} 
\centering
\includegraphics {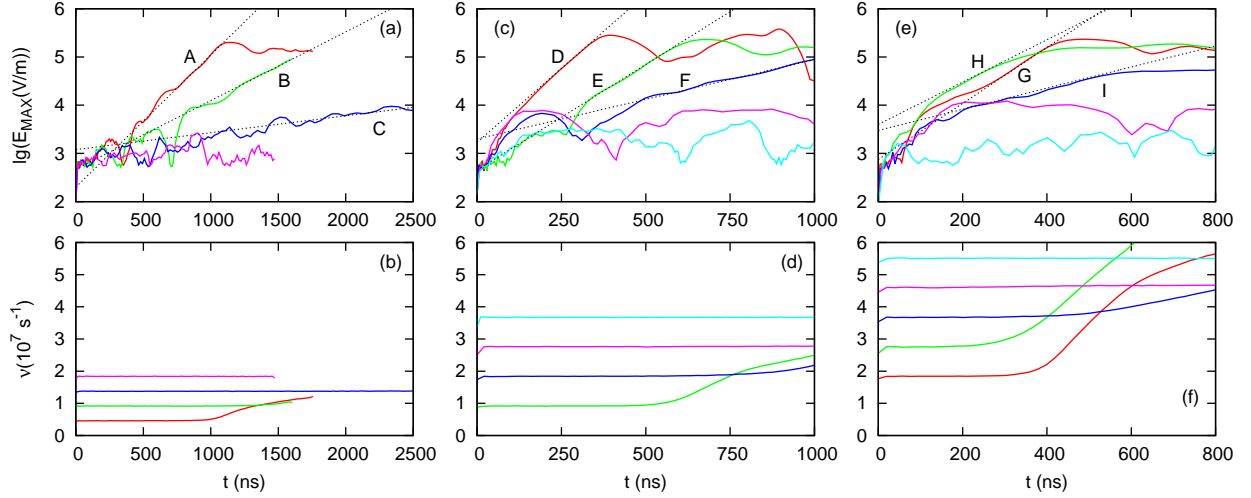}
\caption{\label{fig:pic_logE_nu_vst_3_alpha_L4p5_varcolls_filter} %
The logarithm of amplitude of electric field in V/m (a,c,e) and the bulk
electron collision frequency (b,d,f) versus time in PIC simulations with
parameters shown in Table~\ref{tab:table1}. In (a) and (b), the red, green,
blue, and magenta curves are for simulations 1, 2, 3, and 4 of
Table~\ref{tab:table1}, respectively. In (a), dashed black straight lines A, B,
and C show exponential growth in simulations 1, 2, and 3, respectively. In (c)
and (d), the red, green, blue, magenta, and cyan curves are for simulations 5,
6, 7, 8, and 9, respectively. In (c), dashed black straight lines D, E, and F
show exponential growth in simulations 5, 6, and 7, respectively. In (e) and
(f), the red, green, blue, magenta, and cyan curves are for simulations 10, 11,
12, 13, and 14, respectively. In (e), dashed black straight lines G, H, and I
show exponential growth in simulations 10, 11, and 12, respectively. The growth
rates for lines A to I correspond to the ones shown in Table~\ref{tab:table1}.}
\end{figure*}

In order to obtain a dependence of the threshold beam current on the neutral
gas pressure equivalent to (\ref{eq:26}), a set of fourteen simulations is
carried out with $L=4.5\lambda_b$, three values of electron beam density
$\alpha$, and various values of neutral density $n_n$. The summary of these
simulation parameters is given in Table~\ref{tab:table1}. Note, since the
threshold current (\ref{eq:26}) depends on the neutral pressure $p_n$, below
the pressure is used instead of the density.

The behavior of the system in PIC simulations is qualitatively similar to that
of the fluid system considered above. For each value of $\alpha$, the lower
values of $p_n$ allow development of the instability with a pronounced time
interval of the exponential growth with an approximately constant rate, see
red, green, and blue curves in
Figs.~\ref{fig:pic_logE_nu_vst_3_alpha_L4p5_varcolls_filter}(a), (c), and (e).
Increasing $p_n$ reduces the growth rate and eventually prevents oscillations
from growing after a relatively short initial transitional stage which lasts no
more than 200 ns, see magenta and cyan curves in
Figs.~\ref{fig:pic_logE_nu_vst_3_alpha_L4p5_varcolls_filter}(a), (c), and (e).
The higher the value of $\alpha$, the higher is the value of $p_n$ which
suppresses the instability.

It is necessary to mention that intense oscillations produced by the two stream
instability heat plasma electrons which gradually increases the frequency of
collisions, see the red and the green curves in
Fig.~\ref{fig:pic_logE_nu_vst_3_alpha_L4p5_varcolls_filter}(b), the green and
the blue curves in
Fig.~\ref{fig:pic_logE_nu_vst_3_alpha_L4p5_varcolls_filter}(d), the red, green,
and blue curves in
Fig.~\ref{fig:pic_logE_nu_vst_3_alpha_L4p5_varcolls_filter}(f). The collision
frequency produced by the code diagnostics is the frequency of scattering of
electrons by neutrals averaged over all electron particles. Usually the growth
of the collision frequency is a clear sign that the intense oscillations are
excited. Only in simulation 3 of Table~\ref{tab:table1}, where the oscillations
were growing at a very low rate, no significant modification of the collision
frequency occurs till the end of simulation, see the blue curves in
Figs.~\ref{fig:pic_logE_nu_vst_3_alpha_L4p5_varcolls_filter}(a) and (b). In
simulations where the prolonged exponential growth of oscillations was not
identified, the collision frequency stays constant, see magenta and cyan curves
in Figs.~\ref{fig:pic_logE_nu_vst_3_alpha_L4p5_varcolls_filter}(b), (d), and
(f). Note that the growth of the collision frequency starts when the amplitude
approaches its maximum and does not affect the initial stages of the
instability. The effect of the collisions on the nonlinear stage of the
instability is out of the scope of this paper.

The simulation parameters are selected in such a way that for a single value of
$\alpha$ there are several values of $p_n$. In this case, it is easier to find
the threshold value of the neutral pressure $p_{n,thr}$ preventing the
instability for a given beam current rather then the threshold current for a
given pressure.
To find the threshold pressures, the following procedure is involved.
First, for each simulation where the exponential growth is observed, the growth
rate is identified by fitting the dependence of the electric field amplitude
versus time $E(t)$ with an exponent $\exp(Im(\omega)t)$, see the dashed black
lines A-I in Figs.~\ref{fig:pic_logE_nu_vst_3_alpha_L4p5_varcolls_filter}(a),
(c), and (e). This gives growth rates $Im(\omega)$ for different $\alpha$ and
$p_n$, see Table~\ref{tab:table1}.

\begin{figure}[tbp]
\includegraphics {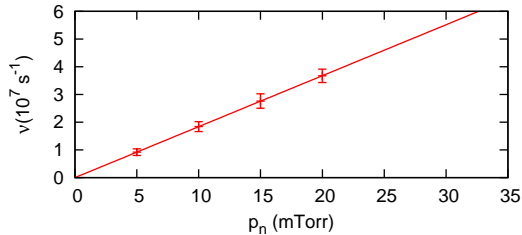}
\caption{\label{fig:pic_nu_vs_pn_L4p5} %
The average frequency of elastic electron-neutral collisions versus the neutral
pressure. Markers with error bars represent values obtained in simulations 6,
7, 8, and 9 of Table~\ref{tab:table1}, the error bars correspond to the
amplitude of noise in the simulation. The solid line is described by equation
(\ref{eq:25}) with $\kappa=1.839\times 10^6 \text{~s}^{-1}\text{mTorr}^{-1}$. }
\end{figure}

\begin{figure}[tbp]
\includegraphics {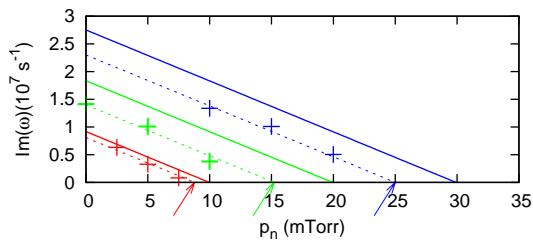}
\caption{\label{fig:Imw_vs_nu_PIC_vs_formula} %
Temporal growth rate of the two-stream instability versus the neutral gas
pressure. The solid straight lines are obtained with fluid theory equations
(\ref{eq:22}) and (\ref{eq:24}) where $\nu_e(p_n)$ is defined by equation
(\ref{eq:25}); the red, green, and blue solid lines are for $\alpha=5\times
10^{-5}$, $10^{-4}$, and $1.5\times 10^{-4}$, respectively. Red markers
represent simulations 1, 2, and 3 of Table~\ref{tab:table1}, respectively (see
also lines A, B, and C in
Fig.~\ref{fig:pic_logE_nu_vst_3_alpha_L4p5_varcolls_filter}(a)). Green markers
represent simulations 5, 6, and 7 of Table~\ref{tab:table1}, respectively (see
also lines D, E, and F in
Fig.~\ref{fig:pic_logE_nu_vst_3_alpha_L4p5_varcolls_filter}(c)). Blue markers
represent simulations 10, 11, and 12 of Table~\ref{tab:table1}, respectively
(see also lines G, H, and I in
Fig.~\ref{fig:pic_logE_nu_vst_3_alpha_L4p5_varcolls_filter}(e)). Dashed red,
green, and blue straight lines are the approximate dependencies ``growth rate
versus pressure'' for $\alpha=5\times 10^{-5}$, $10^{-4}$, and $1.5\times
10^{-4}$, respectively, plotted as the best fit of the PIC simulation data. The
arrows mark threshold pressures which turn the growth rate of the instability into
zero for the three values of the beam density mentioned above. }
\end{figure}

Second, values of $Im(\omega)$ from simulations with the same $\alpha$ but
different $p_n$ are fitted with a straight line
\begin{equation}\label{eq:pic1}
  Im(\omega)=Im(\omega_{ncl}^{kin})-\kappa p_n/2~,
\end{equation}
where $Im(\omega_{ncl}^{kin})$ has the meaning of the growth rate without
collisions in the kinetic description, $\kappa$ is the coefficient of
proportionality between $\nu$ and $p_n$ introduced in Eq.~\ref{eq:25}.
In the present paper, $\kappa=1.839\times 10^6 \text{~s}^{-1}\text{mTorr}^{-1}$
for electrons with a Maxwellian EVDF of temperature 2 eV performing elastic
scattering in an Argon gas with temperature 300 K. This value is obtained by
approximating the initial collision frequencies from simulations 6, 7, 8, and 9
of Table~\ref{tab:table1} with a linear law, as shown in
Fig.~\ref{fig:pic_nu_vs_pn_L4p5}.
Equation (\ref{eq:pic1}) is equivalent to equation (\ref{eq:22}).
Note that while the slope $\kappa/2$ of line (\ref{eq:pic1}) is enforced to
match that of (\ref{eq:22}), the growth rates in PIC simulations fit this slope
surprisingly well, compare markers with dashed straight lines of the same color
in Fig.~\ref{fig:Imw_vs_nu_PIC_vs_formula}.
The difference from the fluid theory is that the collisionless growth rates
$Im(\omega_{ncl}^{kin})$ are lower than the fluid values (\ref{eq:24}) by up to
to $24\%$, compare the dashed and the solid curves of the same color in
Fig.~\ref{fig:Imw_vs_nu_PIC_vs_formula}.
%

Finally, with $Im(\omega_{ncl}^{kin})$ known for each $\alpha$, the threshold
pressure values giving $Im(\omega)=0$ are calculated as
\begin{equation}\label{eq:pic2}
  p_{n,thr}=2Im(\omega_{ncl}^{kin})/\kappa~,
\end{equation}
which is equivalent to (\ref{eq:23}). These threshold pressures are given in
Table~\ref{tab:table1} and are marked by arrows in
Fig.~\ref{fig:Imw_vs_nu_PIC_vs_formula}.
%

\begin{figure}[tbp]
\includegraphics {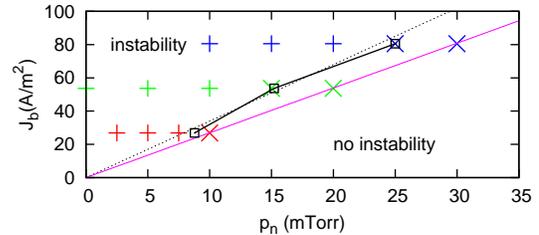}
\caption{\label{fig:pic_jb_vs_pn_L4p5} %
Phase plane ``emission current density vs neutral gas pressure''. Red
markers in ascending pressure order represent simulations 1, 2, 3, and 4 of
Table~\ref{tab:table1}, respectively. Green markers in ascending pressure order
represent simulations 5, 6, 7, 8, and 9, respectively. Blue markers in
ascending pressure order represent simulations 10, 11, 12, 13, and 14,
respectively. Everywhere, the vertical crosses mark simulations where the
exponential growth is identified while the diagonal crosses mark simulations
where the instability is suppressed. Open boxes connected by a solid black line
represent threshold pressures given in Table~\ref{tab:table1} and marked by
arrows in Fig.~\ref{fig:Imw_vs_nu_PIC_vs_formula}. The dashed black straight
line is the threshold current (\ref{eq:pic3}) plotted using these threshold
pressure values. The solid magenta line is the threshold curent (\ref{eq:27})
calculated using the fluid theory. }
\end{figure}

The three values of $p_{n,thr}$ for the three values of beam current (shown by
open black markers connected by a solid black line in
Fig.~\ref{fig:pic_jb_vs_pn_L4p5}) are fitted with a linear law
\begin{equation}\label{eq:pic3}
    J_{b,thr}^{kin}=3.4 p_n
\end{equation}
shown by the black dashed straight line in Fig.~\ref{fig:pic_jb_vs_pn_L4p5}. In
(\ref{eq:pic3}), the current density is in $\text{A}/\text{m}^2$ and the
neutral pressure is in mTorr. This curve represents the threshold current
predicted by the PIC simulation.
For comparison, with $\kappa$, $W_{b,0}$, $n_{e,0}$, and $L_n$ as in the
kinetic simulations with $L=4.5\lambda_b$ above, the fluid threshold current
(\ref{eq:26}) is
\begin{equation}\label{eq:27}
    J_{b,thr}=2.695 p_n~,
\end{equation}
where the current density and the pressure units are the same as in
(\ref{eq:pic3}). Thus, for the same pressure, the value of the threshold
current predicted by PIC simulations is about $26\%$ higher than the value
given by the fluid theory, compare the dashed black and the solid magenta lines
in Fig.~\ref{fig:pic_jb_vs_pn_L4p5}.
This difference corresponds to the lower collisionless growth rates in kinetic
simulation which is reasonable since kinetic effects have a tendency to disrupt
the resonance between the wave and the particles and reduce the growth rate.
Overall, for the selected parameters there is a very reasonable agreement
between the kinetic simulations and the simple fluid theory given in
Section~\ref{sec:02}.
%

\section{Summary \label{sec:05}}

The two-stream instability in a plasma bounded between walls is quite different
from that in an infinite plasma. The oscillations grow both in time and space,
and the growth rates are functions of the distance between the bounding walls.
If this distance is only few resonance wavelengths, the temporal growth rate is
very small. Scattering of plasma bulk electrons further reduces this growth
rate. The present paper finds that the rate becomes zero if the collision
frequency is equal to the doubled growth rate without collisions. Unlike the
results of previous studies,\cite{SinghausPF1964,SelfJAP1971} this criterion
predicts that the instability may be completely suppressed for cold beams.
The proposed fluid theory allows to calculate a threshold beam current density
for the given neutral gas pressure $p_n$, the collision frequency as a function
of the pressure $\nu_e=\kappa p_n$, the electron density $n_e$, the beam energy
$W_b$, and the normalized system length $L_n$:
\begin{equation*}
    J_{b,thr}=\frac{6.5\kappa\sqrt{2e\varepsilon_0 n_e W_b}}
             {L_n\ln(L_n)\left[1-0.18\cos\left(L_n+{\pi}/{2}\right)\right]}
        p_n~,
\end{equation*}
where  $W_b$ is in electronvolts while the units of $p_n$ depend on the method
of calculation of the coefficient $\kappa$.
The instability will not develop if the beam current is below this threshold.
The quantitative effect of collisions on both the growth rate and the threshold
current predicted by the fluid theory is in good agreement with the results of
kinetic simulations.


\section*{ACKNOWLEDGMENTS}

D.~Sydorenko and I.~D.~Kaganovich are supported by the U.S. Department of
Energy. The authors thank A. Khrabrov for his assistance in carrying out the
simulations.

%
%

\end{document}